\documentstyle[stwol]{article}

\def\Journal#1#2#3#4{{#1} {\bf #2}, #3 (#4)}

\def\NPB{{\em Nucl. Phys.} B}
\def\PLB{{\em Phys. Lett.}  B}
\def\PRL{\em Phys. Rev. Lett.}

\def\be{\begin{equation}}
\def\ee{\end{equation}}
\def\bea{\begin{eqnarray}}
\def\eea{\end{eqnarray}}

\def \be  {\begin{equation}}
\def \ee  {\end{equation}}
\def \ba  {\begin{eqnarray}}
\def \ea  {\end{eqnarray}}
\def \baa {\begin{eqnarray*}}
\def \eaa {\end{eqnarray*}}
\newcommand\re[1]{(\ref{#1})}
\newcommand\lr[1]{{\left({#1}\right)}}
\newcommand{\as}{\ifmmode\alpha_{\rm s}\else{$\alpha_{\rm s}$}\fi}
\def \e {\mbox{e}}
\def \ln {\mbox{ln}}
\def \CO {{\cal O}}
\newcommand \ci [1] {\cite{#1}}
\def \lab #1 {\label{#1}}
\newcommand \widebar [1] {\overline{#1}}
\newcommand \vev [1] {\langle{#1}\rangle}
\def\II{\hbox{{1}\kern-.25em\hbox{l}}}

\def \tr {\mbox{tr\,}}
\def \DY {{\rm _{DY}}}
\def \IR {{\rm _{IR}}}

\begin{document}
\begin{titlepage}
\def\thefootnote{\fnsymbol{footnote}}
\thispagestyle{empty}
\hfill\parbox{50mm}{{\sc LPTHE--Orsay--96--78} \par
                         hep-ph/9610207        \par
                         October, 1996}
\vspace*{35mm}
\begin{center}
{\LARGE Power corrections in Drell-Yan production 
\\[2mm]
beyond the leading order}
\par\vspace*{15mm}\par
{\large G.~P.~Korchemsky}
\footnote{E-mail: korchems@qcd.th.u-psud.fr}
\par\bigskip\par\medskip

{\em Laboratoire de Physique Th\'eorique et Hautes Energies
\footnote{Laboratoire associ\'e au Centre National de la Recherche
Scientifique (URA D063)}
\\
Universit\'e de Paris XI, Centre d'Orsay, b\^at. 211\\
91405 Orsay C\'edex, France
\\[3mm]
and
\\[3mm]
Laboratory of Theoretical Physics
\\
Joint Institute for Nuclear Research
\\
141980 Dubna, Russia}

\end{center}
\vspace*{12mm}

\begin{abstract}
We study the power corrections to the cross-section of Drell-Yan
production within the Wilson line approach and apply the methods of QCD
asymptotic dynamics to identify the leading renormalon contribution.
\end{abstract}

\vspace*{35mm}
\begin{center}
{\em To appear in the Proceedings of the \\ 
     28th International Conference on High Energy Physics,\\
     Warsaw, Poland, 25-31 July 1996}
\end{center}

\setcounter{footnote}{0}
\renewcommand{\thefootnote}{\arabic{footnote}}

\end{titlepage}

\title{POWER CORRECTIONS IN DRELL-YAN PRODUCTION \\ BEYOND THE LEADING ORDER}

\author{G.P. KORCHEMSKY}

\address{LPTHE, Universit\'e de Paris XI, 91405 Orsay, France and
LTP, JINR, 141980 Dubna, Russia}

\twocolumn[\maketitle\abstracts{
We study the power corrections to the cross-section of Drell-Yan
production within the Wilson line approach and apply the methods of QCD 
asymptotic dynamics to identify the leading renormalon contribution.
}]

\section{Introduction}

To control the occuracy of perturbative QCD predictions for physical
observables it becomes important to have a regular way of 
estimating corrections suppressed by powers of large momentum scale,
$\Lambda_p/Q^p$, and if not calculate explicitly the nonperturbative 
scale $\Lambda_p$ then find the leading exponent $p$ at least. 
At present, the understanding of power corrections has a solid
theoretical status only for the special class of processes like
$\e^+ \e^-\to\mbox{hadrons}$ and deeply inelastic
scattering (DIS) for which the analysis based on the operator product
expansion (OPE) is applicable. The OPE fixes the structure of power
corrections (as $1/Q^4$ for the total cross-section of $\e^+\e^-$ 
annihilation and $1/Q^2$ for the structure functions of DIS) and 
allows to identify the corresponding scales $\Lambda_p$ as matrix elements 
of higher twist composite operators in QCD. The understanding of power 
corrections in the processes which do not admit the OPE is the subject 
of intensive discussions (see review~\ci{B} and references therein). 

An important example of the process, which does not admit the OPE is the 
Drell-Yan (DY) process $h_A+h_B\to \mu^+\mu^- +X$. Here, the lepton pair
with invariant mass $Q^2$ is created in the partonic subprocess
$q + \bar q\to\mu^+\mu^-(Q^2)+X$ of annihilation of two quarks with 
invariant energy $\hat s$ and we choose the kinematics such that 
$\tau=Q^2/\hat s \sim 1$. One of the reasons for this is that as $\tau\to
1$, the final state in the partonic subprocess consists of the lepton
pair with the energy $Q$ accompanied by a soft gluon radiation with the 
total energy $(1-\tau)Q/2$. Due to the presence of two different scales
the partonic cross-section gets large perturbative (Sudakov logs in 
$\ln(1-\tau)$) and nonperturbative (as inverse powers of $(1-\tau)Q$) 
corrections induced by soft gluons \ci{CS,KS}. 
Calculating the moments $\sigma_N=\int_0^1 d\tau \tau^N d\sigma_{\DY}
/d Q^2$ at large $N\sim\frac1{1-\tau}$ and subtracting collinear 
divergences in the DIS-scheme one can represent the leading term 
of the expansion of $\sigma_N$ in powers of $1/Q$ as 
\be
\ln\,\sigma_N=\sum_k \as^k(Q/N) C_k(\ln N) + \lr{\Lambda_{\DY} N/Q}^p,
\lab{moment}
\ee
where the coupling constant is defined at the characteristic soft 
gluon scale, $C_k$ contains the powers of Sudakov logs plus finite, 
$\ln^0 N$, contributions and the second term represents the leading
power correction. As $N$ increases, the energy of soft gluons in the
final state decreases as $Q/N$ toward the infrared region in which
perturbative expressions should fail. Indeed, it is well-known
that the perturbative expansion in \re{moment} is not well-defined due to
factorial growth of coefficients $C_k \sim \beta_0^k k! p^{-k}$
at higher orders in $\as$. This makes the perturbative series non Borel 
summable and induces an IR renormalon ambiguity into perturbative 
contribution to \re{moment} at the level of power corrections,
$\delta_{\IR} \sigma_N\sim \exp(-\frac{p}{\beta_0\as(Q/N)})
=(\Lambda_{\rm QCD} N/Q)^p$. Then, IR renormalon ambiguity of 
perturbative expression is compensated in \re{moment} by
ambiguity in the definition of the scale $\Lambda_{\DY}$.
Thus, examing the large order behaviour of the perturbative series 
one can determine the level $p$ at which the leading power correction
appear in \re{moment}. However, since it is impossible to calculate 
the coefficients $C_k$ exactly to higher orders in $\as$ one has to 
find a reasonable approximation to $C_k$ which would allow to identify 
the $k!$ factorial growth of the coefficients. To this end different 
schemes were proposed~\ci{BB,DMW,AZ}. Being applied to the Drell-Yan 
cross-section they predict that there are no $N/Q$ corrections to 
\re{moment} and the leading renormalon contribution has the following 
form
\be
\delta_{\IR}\sigma_N = 0\cdot (N/Q) + \lr{A_{\DY} N/Q}^2\,.
\lab{zero}
\ee
We would like to stress that these schemes are {\it approximate\/}
and although they were tested in the case of $\e^+\e^-$ annihilation 
and DIS, it is not clear yet whether they are applicable in the 
Drell-Yan process and whether they really predict the {\it leading\/}
power correction to the cross-section. 

\section{Wilson line approach}

A different approach to the analysis of power corrections has been proposed 
in Ref.~\ci{KS}. It is based on the remarkable property of soft gluons that
their contribution to the Drell-Yan cross-section can be factorized with
a power occuracy into universal factor having the form of a Wilson line.
Similar to the OPE approach, the power corrections can be identified
with the matrix elements of certain operators built from the Wilson line. 

The Wilson line operator appears in the Drell-Yan partonic subprocess as an
eikonal phase of quarks interacting with soft gluon radiation. Combining the
eikonal phases of quark and antiquark together one gets the expression for
partonic cross-section as
$$ 
\frac{d\sigma_{\DY}}{d Q^2}
=\sum_n |\langle n | T U_{\DY}(0) |0\rangle|^2
\delta(E_n-(1-\tau)Q/2)
$$
where summation is performed over all possible final states consisting
of an arbitrary number, $n$, of soft gluons with the total energy
$E_n=(1-\tau)Q/2$ and $T$ stands for time-ordering of gauge fields. 
Here, $U_{\DY}(0)=P\exp(i\int_{C(0)} dx \cdot A(x))$ is the eikonal 
phase of quark + antiquark and
$A_\mu(x)$ is the gauge field operator describing soft gluons.  
The contour $C(0)$ in Minkowski space-time corresponds to the classical
trajectory of quark and antiquark. It goes from $-\infty$ to the
annihilation point $0$ along the quark momentum and then returns 
to $-\infty$ in the direction opposite to the antiquark momentum.
Performing summation over final states in the last relation one can 
obtain the representation for the Drell-Yan cross section for $\tau\to
1$ as a Fourier transformed Wilson line expectation value~\ci{KM}
\ba
\frac{d\sigma_{\DY}}{d Q^2}
&=&
\frac{Q}2\int_{-\infty}^\infty\frac{dy_0}{2\pi}\e^{-iy_0(1-\tau)Q/2}
\langle 0| W(y_0) |0 \rangle
\nonumber
\\
W(y_0)&\equiv&
\widebar T U^\dagger_{\DY}(0)\, T U_{\DY}(y_0)\,,
\lab{W}
\ea
where $U_{\DY}(y_0)$ is defined at the point $y=(y_0,\vec 0)$. 
Calculating the moments of this expression one obtains the following 
relation \footnote{This identity is an analog of the relation between 
the moments of the structure functions of DIS and matrix elements of 
composite operators in the OPE.}
\be
\sigma_N= 
\frac1{N_c} \tr
\langle 0 | W(y_0) |0 \rangle\,,\quad
\mbox{for $y_0=iN/Q$}\,,
\lab{rel}
\ee
which is valid up to corrections vanishing as $N\to \infty$ and which 
takes into account all Sudakov logs and power corrections in $N/Q$. 
The scale $Q/N$ enters into \re{rel} just as a parameter of 
the integration contour in $U_{\DY}(y_0)$. Eq.\re{rel}
states that the asymptotic expansion of
the moments of the Drell-Yan cross section is related to the behaviour
of the Wilson line expectation value at small $y_0=\frac{iN}{Q}$. 

Perturbative expansion of \re{rel} involves standard ``short-distance'' 
logarithms $\as^n\ln^m(|y_0|\mu)$, which are transformed into Sudakov logs
for $y_0=iN/Q$. The factorization scale $\mu\approx Q$ has a meaning of 
the maximal energy of soft gluons contributing to the partonic
cross-section and after factorization of soft gluon emission it
appears in \re{W} and \re{rel} as a {\it ultraviolet\/} renormalization 
scale of the Wilson line operators. As a result, all perturbative Sudakov
corrections to the Drell-Yan cross-section can be effectively resummed 
using the remarkable renormalization properties of Wilson lines. They
are summarized by the following RG equation
\be
\frac{d\ln \sigma_N}{d\ln\mu^2}= 2 \Gamma_{\rm cusp}(\as) \ln(N\mu/Q)
+ \Gamma_{\DY}(\as)\,, 
\lab{RG}
\ee
with $\as=\as(\mu)$ and $\Gamma_{\rm cusp/DY}$ being some anomalous 
dimensions. The evolution of $\sigma_N$ from $\mu=Q/N$ up to $\mu=Q$ 
generates Sudakov logs while the initial condition for $\sigma_N$
at $\mu=N/Q$ takes into account nonperturbative power corrections in
$N/Q$. To understand their structure we perform the operator expansion 
of the Wilson line $W(y_0)$ in powers of $y_0$
\be
\vev{W(y_0)}=\vev{W(0)} + y_0 \vev{\partial W(0)} + ... \ \,.
\lab{ope}
\ee
Substituting this expansion into \re{rel} we find that the power 
corrections to the cross-section are described by matrix elements 
of the operators defined as derivatives of the Wilson line~\ci{KS}. 
Although these operators are nonlocal in QCD, one can represent them as 
local operators in an effective nonabelian Bloch-Nordsieck theory in 
which Wilson line appears as a result of integration over effective
quark fields. Solving the RG 
equation \re{RG} and using \re{ope} as a boundary condition we obtain 
that the leading power correction to the Drell-Yan cross-section appears
as~\ci{KS}
\be
\sigma_N^{\rm power} = \vev{i\partial W(0)}N/Q + \CO(N^2/Q^2)\,,
\lab{fin}
\ee
provided that there is no any hidden symmetry which would prohibit the 
appearence of the linear $\sim y_0$ term in the expansion of the Wilson 
line \re{ope} and would enforce the matrix element
$\vev{i\partial W(0)}$ to vanish. 

\section{IR renormalon analysis}

One possibility to test the presence of the linear term in \re{ope} is to
apply the IR renormalon approach and identify the source of linear
terms inside perturbative series for the Wilson line. At the leading
order of renormalon calculus~\ci{B}, one can perform the explicit one-loop 
calculation of $\vev{W(y_0)}$ with a gluon mass $\lambda$ and identify 
the power corrections as nonanalytical terms in the small $(y_0\,\lambda)$ 
expansion. One finds~\ci{BB} that $\as \ln (y_0\lambda)$ term is absent in 
agreement with the Bloch-Nordsieck cancellation, the linear term 
$\as (\lambda y_0)$ also vanishes and the leading term is 
$\sim\as (\lambda y_0)^2$. This means, that at the leading order the
contribution of IR renormalons has the form \re{zero}.
The same result can be obtained in the 
limit of large number of light quark flavours in which one performs 1-loop 
calculation of $\vev{W(y_0)}$ with the gluon propagator dressed by a 
chain of quark loops~\ci{BB}. We notice that in both cases one calculates 
essentially abelian diagrams containing the color factors $\as C_F$ and 
$\as C_F (\as N_f)^{n}$, respectively. It is well-known that the 
contribution of abelian diagrams to the Wilson line exponentiates to 
higher orders in $\as$ and one can generalize the above statement 
about the absence of linear $\sim y_0$ term to a much larger class of 
abelian diagrams beyond the leading order of the renormalon 
calculus. However, the natural question rises whether the abelian
diagrams provide a meaningful approximation to the exact expression 
for the Wilson line or may be there is the leading contribution
to $W(y_0)$ coming only from nonabelian diagrams. To give an example
where the second possibility is realized we mention the breakdown of
the Bloch-Nordsieck cancellation of IR logs in the Drell-Yan
cross-section at twist $1/Q^4$. The IR divergent part of 
the cross-section has the form~\ci{IR} 
$\sim\as^2 C_A C_F Q^{-4}\ln \lambda^2$ with $C_A (C_F)$ 
being gluon (quark) Casimir operators. It does not appear at order 
$\as$ and at order $\as^2$ it comes only from diagrams with nonabelian 
color structure. One might expect that similar phenomenon may happen when 
one considers linear in $y_0$ contributions to the Wilson line beyond 
the leading order. 

The straighforward way to check this possibility would be to perform 
2-loop calculation of the Wilson line with a power occuracy in
$\lambda$. However, even without going through this complicated
calculation one can employ the operator methods of the QCD asymptotic 
dynamics~\ci{CCM} in order to analyse the properties of the matrix element
\be
\vev{\partial W(0)}=\frac1{N_c}\tr
\vev{
\widebar T U_{\DY}^\dagger(0|A)\,\partial\, T U_{\DY}(0|A)
}\,,
\lab{dW}
\ee
parameterizing the leading $N/Q$ power correction to the Drell-Yan
cross-section. The operator $U_{\DY}(y_0|A)$ describes the eikonal
phase of fast quark and antiquark annihilating at the space-time 
point $y_0$ and $T-$ordering of gauge field operators $A_\mu^a(x)$
is needed in \re{W} and \re{dW} to describe properly the possibility 
for quark and antiquark to exchange by a virtual soft gluon in the
initial state. As a result, due to effects of the initial state 
interaction~\ci{CCM}
$$
TU_{\DY}(y_0|A)\neq U_{\DY}(y_0|A)\,.
$$
This does not happen however for single-parton initiated process like 
DIS and it is this phenomenon that makes different the properties of 
power corrections in DY and DIS. Indeed, if we omit the $T-$ordering 
in \re{dW} then it is easy to show by direct calculation that the derivative
$U_{\DY}^\dagger\partial U_{\DY}$ is proportional to the 
generators of the $SU(N_c)$ group and it does not contribute to 
\re{dW}. To work out the effects of quark-antiquark correlations in
$\vev{W(y_0)}$ one defines two gauge field operators
$A_\mu^\pm(x)=\frac12(A_\mu(x)\pm A_\mu(-x))$ and notices~\ci{CCM}
that the field $A_\mu^+$ describes the Coulomb gluons while the field 
$A_\mu^-$ is associated with radiative gluons forming quark and
antiquark coherent states. Since $[A^{a,\pm}_\mu(x),A^{b,\pm}_\nu(y)]=0$ and
$[A^{a,\pm}_\mu(x),A^{b,\mp}_\nu(y)]\neq 0$, the expansion of the matrix
element $\vev{n|TU_{\DY}(y_0|A^++A^-)|0}$ takes the following form~\ci{CCM}
\ba
&&\hspace*{-5mm}
\vev{n|TU_{\DY}^{ij}
(y_0|A^++A^-)|0}
=[\e^{i\Phi_C}]^{ij}_{i'j'}
\lab{V}
\\
&&\hspace*{5mm}
\times
\vev{n|
U_{\DY}^{i''j''}(y_0|A^-)
V_{i''j''}^{i'j'}(y_0|A^+,A^-)
|0} \,,
\nonumber
\ea
where the flow of quark color is indicated explicitly.
Here, in the r.h.s.\ the first factor is the nonabelian unitary
Coulomb phase matrix and the eikonal phase $U_{\DY}$ depends
only on commutative operators $A^-$. The residual
factor $V$ is unitary, $V^\dagger V=\II$, and it
describes the correlations between Coulomb and radiative
gluons. The perturbative expansion of $V$ looks like~\ci{CCM}
$$
V=\II+ig^3 f^{abc}t^a\otimes t^b\!\int\!\! d^4x\, 
G(x,y_0) A_\mu^{c,+}(x)+\CO(g^4)
$$
where effective nonlocal coupling $G(x,y_0)$ resembles the 
Fadin-Kuraev-Lipatov vertex. In abelian theory $V=\II$ to all orders, 
but in QCD it gets {\it nonabelian} corrections starting from order 
$g^3$ in the strong coupling constant. 

Substituting \re{V} into \re{dW} one observes that 
nonabelian Coulomb phase is canceled. Moreover, if we neglect 
$g^3-$corrections to $V$, then due to unitarity and commutativity of 
$U_{\DY}(y_0|A^-)$, the operator $U_{\DY}^\dagger\partial U_{\DY}$
is again proportional to the $SU(N_c)$ generators and its matrix element
vanishes identically in Eq.\re{dW}. Thus, a nonzero contribution to 
\re{dW} comes only from higher order terms in $V$. It immediately 
follows from the explicit form of $V$ that the corresponding Feynman 
diagrams should have a nonabelian color structure and they arise 
starting from $\as^2$ order. The set of relevant 2-loop Feynman 
diagrams was identified in Ref.~\ci{CCM}. This also explains why 
linear term did not appear in 1-loop calculation of the Wilson line 
performed in Ref.~\ci{BB}.

The fact that the contribution of the residue factor $V$ to the Wilson 
line $W(y_0)$ is nonzero was demonstrated by a direct calculation 
in Ref.~\ci{CCM}, in which a small gluon mass $\lambda$ was used as an 
IR cutoff and only logarithmic terms, $\sim\ln^k\lambda$, were kept. 
It was shown that the leading term $W(0)$ of the expansion of the
Wilson line in Eq.\re{ope} gets a higher twist correction 
$\sim\as^2 C_FC_A Q^{-4}\ln\lambda$, which leads to the breakdown of 
Bloch-Nordsieck theorem for the Drell-Yan cross-section but is 
nevertheless consistent with the Kinoshita-Lee-Nauenberg (KLN) theorem. 
To find the linear term in \re{ope} one has to perform the
calculation of the same diagrams with a power 
accuracy in $\lambda$, although interpretation of the linear term
$\sim\lambda$ to order $\as^2$ as a ``true'' power correction is not 
obvious. One may dress instead the gluon propagators by a single 
chain of quark loops, resum all 2-loop diagrams of order 
$\as^2 C_FC_A(\as N_f)^k$ in the large $N_f$ limit and identify the 
contribution of the leading IR renormalon.  

\section{Conclusions}

We conclude that the effects of the initial state interaction, that is
correlations between Coulomb and radiative gluons, provide a new source 
of $N/Q$ power corrections in the Drell-Yan process. In the Wilson
line approach they are described by the matrix element of the  
derivative of the Wilson line operator in Eq.\re{fin}. In the IR
renormalon analysis these effects could manifest themselves only  
in nonabelian Feynman diagrams at higher orders in $\as$.
The presence of $N/Q$ corrections in the Drell-Yan cross-section is
consistent with the KLN theorem. Summation over degenerate initial 
partonic states in \re{W} and \re{rel} 
leads to the expression for the KLN cross-section, $\sigma_N^{\rm _{KLN}}=
\sum_k \vev{k|W(iN/Q)|k}$, where sum goes over an arbitrary number of
incoming soft gluons. As was shown in Ref.~\ci{CCM}, the initial 
state interaction effects can be analysed in $\sigma_N^{\rm _{KLN}}$ on 
the same footing as the final state effects thus resulting in the 
cancellation of $N/Q$ corrections in $\sigma_N^{\rm _{KLN}}$.

\section*{Acknowledgements}

The author is most grateful to George Sterman for useful discussions.

\end{document}